\documentclass[12pt,preprint]{aastex}
\usepackage{amsmath}
\usepackage{lscape}

\shorttitle{{Gamma-ray emission from the globular clusters}}
\shortauthors{Tam et al.}
\begin{document}

\title{Gamma-ray emission from the globular clusters Liller~1, M80, NGC~6139, NGC~6541, NGC~6624, and NGC~6752}

\author{P. H. T. Tam$^{1}$, A. K. H. Kong$^{1,4}$, C. Y. Hui$^{2}$, K. S. Cheng$^{3}$, C. Li$^{3}$, and T.-N. Lu$^{1}$}
\affil {$^1$ Institute of Astronomy and Department of Physics, National Tsing Hua University, Hsinchu, Taiwan\\
$^2$ Department of Astronomy and Space Science, Chungnam National University, Daejeon, South Korea\\
$^3$ Department of Physics, University of Hong Kong, Pokfulam Road, Hong Kong\\
$^4$ Golden Jade Fellow of Kenda Foundation, Taiwan}
\email{phtam@phys.nthu.edu.tw}

\begin{abstract}
Globular clusters (GCs) are emerging as a new class of $\gamma$-ray emitters, thanks to the data obtained from the Fermi Gamma-ray Space Telescope. By now, eight GCs are known to emit $\gamma$-rays at energies $>$100~MeV. Based on the stellar encounter rate of the GCs, we identify potential $\gamma$-ray emitting GCs out of all known GCs that have not been studied in details before. In this paper, we report the discovery of a number of new $\gamma$-ray GCs: Liller~1, NGC~6624, and NGC~6752, and evidence for $\gamma$-ray emission from M80, NGC~6139, and NGC~6541, in which $\gamma$-rays were found within the GC tidal radius. With one of the highest metallicity among all GCs in the Milky Way, the $\gamma$-ray luminosity of Liller~1 is found to be the highest of all known $\gamma$-ray GCs. In addition, we confirm a previous report of significant $\gamma$-ray emitting region next to NGC~6441. We briefly discuss the observed offset of $\gamma$-rays from some GC cores. The increasing number of known $\gamma$-ray GCs at distances out to $\sim$10~kpc is important for us to understand the $\gamma$-ray emitting mechanism and provides an alternative probe to the underlying millisecond pulsar populations of the GCs.
\end{abstract}


\section{Introduction}

Globular clusters (GCs) are very efficient in forming close binary systems and their decedents such as low-mass X-ray binaries (LMXBs), cataclysmic variables, and millisecond pulsars (MSPs). It is commonly believed that the high stellar encounter rate of GCs facilitates the formation of these binary systems through dynamical interactions~\citep[e.g.,][]{Verbunt87}. \citet{Pooley03} and \citet{Pooley06} have shown by X-ray observations that the number of close binary systems in a GC scales with the stellar encounter rate ($\Gamma_\mathrm{c}$) of the GC. Recently, \citet{Hui10_metallicity} found that a cluster with higher $\Gamma_{c}$ and metallicity hosts more MSPs.

Radio and X-ray observations have revealed about 140 MSPs in 26 globular clusters\footnote{http://www.naic.edu/\textasciitilde pfreire/GCpsr.html}~\citep{atnf_cat}. However, the presence of much stronger X-ray emitters can contaminate the X-ray observations of MSPs. \citet{lat_8_field_GCs} reported a number of MSPs in the Galactic field using data obtained using the Fermi Gamma-ray telescope. Because MSPs are the only known steady $\gamma$-ray sources in GCs, $\gamma$-ray observations serve as an alternative channel in studying the underlying MSP populations in GCs.

The Large Area Telescope (LAT), onboard the \emph{Fermi Gamma-ray Space Telescope}, is a pair-production telescope designed to detect $\gamma$-rays with energies between $\sim$20~MeV and $>$300~GeV. It operates in a survey mode in which it scans the whole sky every 3 hours. The 68\% containment radius of individual photons is $0\fdg6$ at 1~GeV, the typical energy of the photons collected from the GCs. The point-source sensitivity of LAT is $\sim$10$^{-8}\,\mathrm{ph}\,\mathrm{cm}^{-2}\,\mathrm{s}^{-1}$ above 100~MeV in one year of survey-mode observations~\citep{lat_technical}. Such sensitivity has enabled the discovery of $\gamma$-rays from 8 GCs~\citep{lat_1st_cat,lat_13_GCs}, including 47~Tucanae~\citep{lat_47Tuc_Science} and Terzan~5~\citep{Kong10_Terzan5}. From the known $\gamma$-ray luminosity of individual MSPs in the Galactic field~\citep{lat_8_field_GCs}, they would not be detected at distances of several kpc. It is therefore believed that the $\gamma$-rays from GCs do not come from individual MSP, but from a population of MSPs.

The radiation mechanism of $\gamma$-rays remains unclear. In the pulsar magnetosphere model~\citep[e.g.,][]{Venter08}, $\gamma$-rays up to a few GeV is radiated from the MSPs through curvature radiation. On the other hand, inverse Compton (IC) processes resulted from energetic particles up-scattering low-energy photons, such as starlight and infrared light, may give rise to $\gamma$-rays of MeV to TeV energies~\citep{Bednarek07,Cheng_ic10}. It is worth noting that $\gamma$-ray observations at $>$100 GeV have not yet resulted in any detections~\citep{magic_M13,hess_47Tuc}.

In this work, we searched for $\gamma$-rays from those GCs with high encounter rate. We report the discovery of $\gamma$-rays from the GCs Liller~1, M80, NGC~6139, NGC~6541, and NGC~6752.

\section{Cluster sample and Fermi Observations}
\label{sect:obs}

It has long been suggested that the stellar encounter rate of a GC is a measure of the number of LMXBs, the precedents of MSPs~\citep[see, e.g.][]{Verbunt87,Gendre2003}. Assuming that each MSP emits similar amount of $\gamma$-rays, the expected $\gamma$-ray luminosity of a GC is proportional to the number of MSPs it hosts, and thus to the encounter rate $\Gamma_\mathrm{c}$, as supported by the findings made by \citet{lat_13_GCs}. Here we estimate the encounter rate by $\Gamma_\mathrm{c}\propto\rho_0^2\,r_\mathrm{c}^3/\sigma$ where $\rho_0$ is the central luminosity density, $r_\mathrm{c}$ the core radius, and $\sigma$ the velocity dispersion at the cluster center. We adopted the values of $\sigma$ as presented in~\citet{Gnedin02}, and those of $\rho_0$ and $r_\mathrm{c}$ in~\citet[][2003 version]{harris_catalog}. Then we ranked the GCs according to the encounter rate divided by the squared distance compiled in~\citet[][2003 version]{harris_catalog}, giving a measure of the relative expected $\gamma$-ray flux for each GC. Most of the eight known $\gamma$-ray GCs (47~Tucanae, $\omega$ Centauri, M62, NGC~6388, Terzan~5, NGC~6440, M28, and NGC~6652) and three source candidates (NGC~6541, NGC~6752, and M15) are ranked high, i.e. at the top 23 in the ranking. Only NGC~6652 gave a relatively low expected $\gamma$-ray flux. Among the top 20, we identify the following GCs that have not been reported as $\gamma$-ray source or source candidate before (in descending order of expected $\gamma$-ray flux): Liller~1, M22, NGC~2808, NGC~362, NGC~6540, NGC~1851, Terzan~6, M80, and NGC~6397. On the other hand, nearby $\gamma$-ray emission has been reported for NGC~6441 and NGC~6624~\citep{lat_13_GCs}. 
In this work, we searched for $\gamma$-rays from the GCs in the above list, as well as looked deeper into the cases of NGC~6441, NGC~6541, NGC~6624, and NGC~6752. NGC~6139 (ranked 30th) was also studied.

The $\gamma$-ray data used in this work were obtained between 2008 August 4 and 2010 August 21. We used the Fermi Science Tools v9r15p2 package to reduce and analyze the data provided by the Fermi Science Support Center\footnote{http://fermi.gsfc.nasa.gov/ssc/data/analysis/scitools/}. Only those data that passed the most stringent photon selection criteria (i.e. the ``diffuse'' class) were used. To reduce the contamination from Earth albedo
$\gamma$-rays, we excluded events with zenith angles greater than 105$\degr$. The instrument response functions (IRFs)
``P6\_V3\_DIFFUSE'' were used.

\section{data analysis and results}

Photons with energies between $E_\mathrm{min}$ and $E_\mathrm{max}$ that come from a circular region-of-interest (ROI) around the GCs were included in the analysis. The ROI, as well as $E_\mathrm{min}$ and $E_\mathrm{max}$, for those GC coincident with $\gamma$-ray emission are given in Table~\ref{GC_data}. We subtracted the background contribution from all sources in the first Fermi/LAT catalog~\citep[1FGL;][]{lat_1st_cat} within the circular region of 15$^\circ$ radius around the GC position, as well as diffuse emission by including Galactic diffuse model (gll\_iem\_v02.fit) and isotropic background (isotropic\_iem\_v02.txt). We assumed single power laws for all first Fermi/LAT catalog sources considered, except for $\gamma$-ray pulsars of which the spectra follow power laws with exponential cut-off~\citep{lat_1st_psr_cat,lat_13_GCs}.

We created the ``Test-statistic~\citep[TS;][]{Mattox_96} maps'' in the neighboring region of the clusters using the tool \emph{gttsmap}. These TS maps are created by moving a putative point source through a grid of locations on the sky and maximizing $-$log(likelihood) at each grid point, while stronger and presumably well-identified sources are included in each fit. Weaker sources would be identified in these maps. Such TS maps around Liller~1, M80, NGC~6139, NGC~6441, NGC~6541,
NGC~6624, and NGC~6752 are shown in Figs.~\ref{Liller1_M80_TSmap} to \ref{6441_TSmap} centered on the best-fit centroid of the $\gamma$-ray emission (to be determined in the following). Photons with energy between $E_\mathrm{min}$ and $E_\mathrm{max}$ as stated in Table~\ref{GC_data} were used in generating the maps, except for Liller~1 whose map was generated using photons with energy between 2--30~GeV to avoid contamination from the strong Galactic background due to its Galactic position of $l=354\fdg84$ and $b=-0\fdg16$.

We found evidence for $\gamma$-ray emission within the tidal radius of Liller~1, M80, NGC~6139, NGC~6541, NGC~6624, and NGC~6752. In addition, we found significant $\gamma$-ray emission next to NGC~6441, previously reported in~\citet{lat_13_GCs}. In all these cases, a $\gamma$-ray source candidate in the neighborhood of the corresponding GCs is detected above the modeled background emission. We did not find significant $\gamma$-ray emission within the tidal radius of other candidates: M22, NGC~2808, NGC~362, NGC~6540, NGC~1851, Terzan~6, and NGC~6397.

The TS values and the corresponding significances ($=\sqrt{TS}$) of the $\gamma$-ray emission coincident with the GCs are shown in Table~\ref{GC_results}. The number of gamma-ray photons associated with the corresponding $\gamma$-ray source in the best-fit model provided by the likelihood tool \emph{gtlike} are also listed. Since the significance values for a number of positions were calculated, this significance has to be corrected according to a ``trial factor'' to give a ``post-trial'' significance. This trial factor accounts for the increased probability of finding a fake signal with a number of search positions. We estimate the trial factor as $N_\mathrm{GC}\times N_\mathrm{bin}$, where $N_\mathrm{GC}$ is the number of GCs searched, and $N_\mathrm{bin}$ the number of bins within the tidal radius of each GC. We have searched for $\gamma$-rays from 30 GCs\citep[including those with the highest encounter rate normalized by its distance and those first mentioned in][]{lat_13_GCs}, therefore $N_\mathrm{GC}=30$. For $N_\mathrm{bin}$, the number of bins is obtained by dividing the area of the tidal radius of an GC (we take the tidal radius of Liller~1, 13$\arcmin$, as the averaged tidal radius) by the area of each bin ($0\fdg1\times0\fdg1=0.01$~squared degrees). This gives $N_\mathrm{bin}\sim15$. After accounting for the trial factor, the ``post-trial'' detection significance values are obtained. Since the TS value for adjacent grid points are correlated in the likelihood analysis procedure, the post-trial significance can be considered as conservative estimates.

We then used \emph{gtfindsrc} to determine the positions of the $\gamma$-ray source candidates and \emph{gtlike} (unbinned likelihood analysis) to obtain their spectra and their TS values. The integrated photon fluxes and energy fluxes are given for the energy range 100~MeV to 100~GeV, using simple extrapolation from the respective models. For the three cases of most significant detection (i.e., Liller~1, NGC~6624, and the $\gamma$-ray source next to NGC~6441), photons between 300~MeV and 100~GeV are divided into 6 logarithmically-equally spaced energy bins. The flux in each bin was obtained by \emph{gtlike} using photons in the corresponding bin, while both the normalization and photon indices of all $\gamma$-ray sources and the diffuse background components, namely Galactic diffuse and isotropic components, were set to be free. Only for cases where the photon statistics is not enough for a satisfactory fit, the photon index was fixed at the best-fit value shown in Table~\ref{GC_results}.

We also performed a long-term temporal analysis of the 7 $\gamma$-ray sources, in which the 2-year data were binned into several periods depending on the photon statistics. No significant $\gamma$-ray variability nor flaring period was found, indicating that the sources are stable in radiating $\gamma$-rays.

In the following section, we discuss the results of each of the GCs in more details.

\section{Gamma-ray properties from individual clusters}

\subsection{Gamma-ray emission within cluster's tidal radius}

\subsubsection{Liller~1}

The encounter rate of Liller~1 is the highest among the target list mentioned in Section~\ref{sect:obs}.
As seen in Fig.~\ref{Liller1_M80_TSmap}, $\gamma$-ray emission from the direction of Liller~1 was found. Fitting the photon spectrum ($dN/dE$) with the single power law, the TS value of this source is 107, corresponding to a detection significance of 10.3$\sigma$ (i.e. 9.4$\sigma$ post-trial). The best-fit position is R.A.~$=263\fdg20$, Dec.~$=-33\fdg39$ $\pm 0\fdg04$(stat) (J2000; 1-$\sigma$ uncertainty). For localization of a typical $\gamma$-ray source, the systematic position error is estimated to be about 40\% of the statistical error~\citep{bsl_lat}\footnote{We assume similar systematic errors for other GCs as well.}. The best-fit position is $7\farcm5$ from the core position of the GC Liller~1, which is within its tidal radius ($12\farcm57$).
Fig.~\ref{Liller1_spec} depicts the spectrum (in $E^2 dN/dE$) of the $\gamma$-ray emission between 300~MeV and 100~GeV. The photon spectrum of Liller~1 can be well fit by a single power law with an index of $2.2\pm0.1$, and the integral photon flux between 100~MeV and 100~GeV is
$F_{0.1-100\,\mathrm{GeV}}=(6.9\pm2.3)\times10^{-8}\mathrm{cm}^{-2}\,\mathrm{s}^{-1}$ in this model.
On the other hand, the parameters provided by the power law with an exponential cut-off (PLE) model with all three parameters being free are not well constrained; we then fixed the cut-off energy in this model at 2~GeV and 10~GeV, and obtained smaller values of TS (i.e. 90 and 100, respectively). We therefore do not consider this model to be significant. We found that the rightmost spectral point shown in Fig.~\ref{Liller1_spec}, that corresponds to 40--100~GeV emission, were detected with a TS value of 16, corresponding to a significance of 4$\sigma$. At a distance of 9.6~kpc \citep[][2003 version; which is within the range (8.3$\pm$1.8)~kpc given in~\citet{Ortolani_Liller1_UKS1_dist}]{harris_catalog}, the $\gamma$-ray luminosity of Liller~1 is $L_{0.1-100\,\mathrm{GeV}}=(5.9\pm2.0)\times10^{35}\mathrm{erg}\,\mathrm{s}^{-1}$, which is the largest among all detected $\gamma$-ray emitting GCs by now.

\subsubsection{M80}

Among the GCs in the target list, M80 was found to be a possibly $\gamma$-ray emitting GC.
This GC is located at the Galactic coordinate of $(l,b)=(352\fdg6\mathrm{,}19\fdg4)$, i.e. well outside the Galactic plane. Therefore we extended the ROI radius to $10\degr$ and included photons with energies 200~MeV to 100~GeV in the analysis. Inside the tidal radius of M80 we found evidence for $\gamma$-ray emission that peaks at a TS value of 27, corresponding to a significance of 5.2$\sigma$ (i.e. 3.9$\sigma$ post-trial), as shown in Fig.~\ref{Liller1_M80_TSmap} (region \emph{A}). We therefore regard M80 as a possible $\gamma$-ray emitter. In addition, it is possible that the possible $\gamma$-ray emission region may extend well outside the GC, i.e. into region \emph{B}. Formally we cannot separate the whole $\gamma$-ray emission into two emitting regions. Assuming that the $\gamma$-ray emitting region coincident with the GC (i.e. region \emph{A}) is a separate source, its best-fit position is R.A.~$=244\fdg23$, Dec.~$=-23\fdg02$ $\pm 0\fdg06$(stat) (J2000). Its distance of $3\farcm3$ from the GC core position makes the two positions consistent within errors. The photon spectrum of this $\gamma$-ray source can be well fit by a single power law with an index of $2.0\pm0.2$. The integral $\gamma$-ray flux is 
$F_{0.1-100\,\mathrm{GeV}}=(5.5\pm3.6)\times10^{-9}\mathrm{cm}^{-2}\,\mathrm{s}^{-1}$. The low statistics of photons does not allow for accurate determination of the parameter values in the PLE model; we therefore do not consider this model. At a distance of $10.3^{+0.8}_{-0.7}$~kpc~\citep{brocato_M80_dist}, $L_{0.1-100\,\mathrm{GeV}}=8.4^{+7.6}_{-5.8}\times10^{34}\mathrm{erg}\,\mathrm{s}^{-1}$.

The best-fit position of the $\gamma$-ray emitting region \emph{B} (TS$=$30) is R.A.~$=244\fdg42$, Dec.~$=-23\fdg50$ $\pm 0\fdg07$(stat) (J2000). The photon spectrum of this $\gamma$-ray source can be fit by a single power law with an index of $2.2\pm0.2$. The integral $\gamma$-ray flux is $F_{0.1-100\,\mathrm{GeV}}=(1.1\pm0.6)\times10^{-8}\mathrm{cm}^{-2}\,\mathrm{s}^{-1}$.

\subsubsection{NGC~6139}

The encounter rate of NGC~6139 is the 30th highest in our target list. We found evidence for $\gamma$-ray emission at the position of NGC~6139. The TS value of this source candidate is 32, corresponding to a detection significance of 5.6$\sigma$ (i.e. 4.5$\sigma$ post-trial), obtained in the power law model. We therefore regard this detection as a marginal detection. 
The TS map around NGC~6139 is shown in Fig.~\ref{6139_6541_TSmap}.
The best-fit position is R.A.~$=246\fdg83$, Dec.~$=-38\fdg90$ $\pm 0\fdg07$(stat) (J2000), which is $5\farcm1$ from the core position of NGC~6139, but within its tidal radius ($8\farcm52$). If the $\gamma$-ray source is confirmed, the $\gamma$-ray spectrum of NGC~6139 can be well fit by a single power law with an index of $2.1\pm0.2$. The integrated photon flux between 100~MeV and 100~GeV is
$F_{0.1-100\,\mathrm{GeV}}=(9.9\pm5.4)\times10^{-9}\mathrm{cm}^{-2}\,\mathrm{s}^{-1}$.
The low statistics of photons does not allow for accurate determination of the parameter values in the PLE model; we therefore do not consider this model. At a distance of 10.1~kpc~\citep[][2003 version]{harris_catalog}, the $\gamma$-ray luminosity of NGC~6139 is $L_{0.1-100\,\mathrm{GeV}}=(1.1\pm0.6)\times10^{35}\mathrm{erg}\,\mathrm{s}^{-1}$.

\subsubsection{NGC~6541}

Inside the tidal radius of NGC~6541, we found a possible $\gamma$-ray source that peaks at TS$=$19, corresponding to a significance of 4.4$\sigma$ (i.e. 2.8$\sigma$ post-trial). Therefore, we consider this source to be possibly detected, consistent with the finding of~\citet{lat_13_GCs}. The TS map is shown in Fig.~\ref{6139_6541_TSmap}. The best-fit position is R.A.~$=272\fdg04$, Dec.~$=-43\fdg85$ $\pm 0\fdg11$(stat) (J2000), $8\farcm9$ from the core position of the GC NGC~6541, well within its tidal radius ($29\farcm60$). This position is consistent with those derived in~\citet{lat_1st_cat,lat_13_GCs}. Therefore, we identify this source candidate to be 1FGL~J1807.6$-$4341. A power-law fit of the photon spectrum gave an index of $2.2\pm0.2$ and an integrated photon flux of
$F_{0.1-100\,\mathrm{GeV}}=(9.5\pm4.4)\times10^{-9}\mathrm{cm}^{-2}\,\mathrm{s}^{-1}$. This flux is consistent with the upper limit of $1.1\times10^{-8}\mathrm{cm}^{-2}\,\mathrm{s}^{-1}$ as given in~\citet{lat_13_GCs}, noting that their flux was calculated using the PLE model. Moreover, our derived integrated energy flux of $E_{0.1-100\,\mathrm{GeV}}=(6.5\pm3.0)\times10^{-12}\mathrm{erg}\,\mathrm{cm}^{-1}\,\mathrm{s}^{-1}$ are consistent with that given in \citet{lat_1st_cat}.
The PLE model was not used in the likelihood fit because of the putative nature of the $\gamma$-ray emission. At a distance of ($6.9\pm0.7$)~kpc~\citep{LeeCarney06}, the $\gamma$-ray luminosity of the $\gamma$-ray source candidate is $L_{0.1-100\,\mathrm{GeV}}=3.7^{+2.9}_{-2.1}\times10^{34}\mathrm{erg}\,\mathrm{s}^{-1}$.

\subsubsection{NGC 6624}

The $\gamma$-ray source 1FGL~J1823.4-3009 lies close to NGC~6624~\citep{lat_1st_cat,lat_13_GCs}. Using a larger data set, we confirmed this $\gamma$-ray source and obtained a TS value of 121, corresponding to a detection significance of 11.0$\sigma$ (i.e. 10.1$\sigma$ post-trial). The best-fit position was found to be R.A.~$=275\fdg86$, Dec.~$=-30\fdg16$ $\pm 0\fdg05$(stat) (J2000), $12\farcm2$ from the core position of NGC~6624 (shown in Fig.~\ref{6624_6752_TSmap}). This position is consistent with that reported in~\citet{lat_13_GCs} and it lies within the tidal radius of the cluster ($20\farcm55$). We obtained a single power-law fit of the photon spectrum with an index $2.2\pm0.1$, in which the integrated $\gamma$-ray flux is $F_{0.1-100\,\mathrm{GeV}}=(2.6\pm0.6)\times10^{-8}\mathrm{cm}^{-2}\,\mathrm{s}^{-1}$. The above power-law index and our derived integrated energy flux of $E_{0.1-100\,\mathrm{GeV}}=(2.0\pm0.5)\times10^{-11}\mathrm{erg}\,\mathrm{cm}^{-1}\,\mathrm{s}^{-1}$ are consistent with that given in \citet{lat_1st_cat}. On the other hand, the parameters provided by the PLE model (with all three parameters free) are not well constrained; we then fixed the cut-off energy in this model at 600~MeV, 1~GeV, 1.5~GeV, 2~GeV, 3~GeV and 5~GeV, and obtained TS values of 111, 120, 121, 120, 118, and 116, respectively). Fixing the cutoff energy at 1.5~GeV, the power law index was found to be $0.9\pm0.2$, and the integrated $\gamma$-ray flux was found to be $F_{0.1-100\,\mathrm{GeV}}=1.0^{+0.7}_{-0.5}\times10^{-8}\mathrm{cm}^{-2}\,\mathrm{s}^{-1}$. Since this model gives the same TS($=121$) as the power law, it is a model as good as the power-law model.
We found that the two rightmost spectral points shown in Fig.~\ref{ngc6624_spec}, that together correspond to 14--100~GeV emission, were detected with a TS value of $\sim$10 each, corresponding to a significance of 3$\sigma$ each. At a distance of 7.9~kpc~\citep[][2003 version]{harris_catalog}, the $\gamma$-ray luminosity of NGC~6624 (using the power-law model) is $L_{0.1-100\,\mathrm{GeV}}=(1.5\pm0.3)\times10^{35}\mathrm{erg}\,\mathrm{s}^{-1}$.

\subsubsection{NGC~6752}

We found significant $\gamma$-ray emission from NGC~6752. This source has a TS value of 49, corresponding to a significance of 7.0$\sigma$ (i.e. 6.0$\sigma$ post-trial), in the power law model. As shown in Fig.~\ref{6624_6752_TSmap}, the source is compact and located close to the core of NGC~6752. The best-fit position is R.A.~$=287\fdg57$, Dec.~$=-59\fdg96$ $\pm 0\fdg09$(stat) (J2000), $4\farcm7$ from the core position of NGC~6752, and is consistent with being emitted from within the half-mass radius, as shown in Fig.~\ref{6624_6752_TSmap}. This position is consistent with that derived in~\citet{lat_13_GCs}. The photon spectrum of the source can be well fit by a single power law with an index of $2.0\pm0.2$. The integrated photon flux between 100~MeV and 100~GeV is
$F_{0.1-100\,\mathrm{GeV}}=(6.0\pm2.8)\times10^{-9}\mathrm{cm}^{-2}\,\mathrm{s}^{-1}$. This flux is consistent with the upper limit of $7\times10^{-9}\mathrm{cm}^{-2}\,\mathrm{s}^{-1}$ as given in~\citet{lat_13_GCs}, noting that their flux was calculated using the PLE model.
The parameters provided by the PLE model (with all three parameters free) are not well constrained; we then fixed the cut-off energy at 600~MeV, 1~GeV, 3~GeV and 10~GeV, and obtained smaller values of TS (i.e. 23, 27, 34, and 42, respectively). We therefore do not consider this model to be significant. At a distance of ($4.4\pm0.1$)~kpc~\citep[][2003 version]{harris_catalog}, the $\gamma$-ray luminosity of NGC~6752 is $L_{0.1-100\,\mathrm{GeV}}=(1.4\pm0.7)\times10^{34}\mathrm{erg}\,\mathrm{s}^{-1}$.

\subsection{Gamma-ray emission outside cluster's tidal radius: the case of NGC 6441}

\begin{figure}
    \epsscale{1.2}
    \plotone{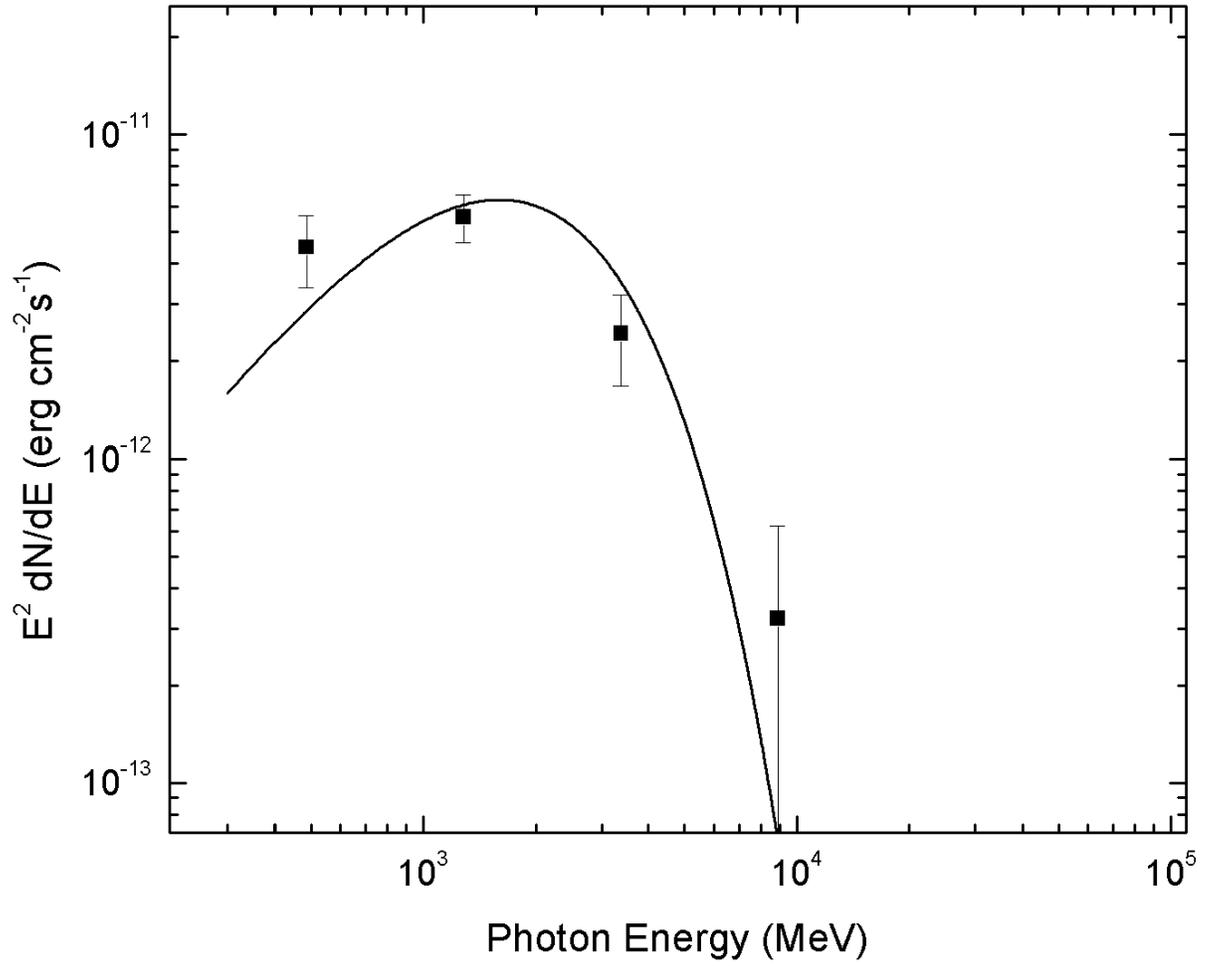}
      \caption{The spectrum ($E^2\times dN/dE$) of the $\gamma$-ray emission outside the tidal radius of NGC~6441. The solid line indicates the best-fit PLE model.}
         \label{ngc6441_spec}
   \end{figure}

A $\gamma$-ray source next to NGC~6441 was first reported by~\citet{lat_13_GCs}. Using a larger data set, we also found a source (a TS value of 101 is obtained with the PLE model, corresponding to a detection significance of 10.0$\sigma$, i.e. 9.1$\sigma$ post-trial) with the best-fit position of R.A.~$=267\fdg63$, Dec.~$=-36\fdg89$ $\pm 0\fdg09$(stat) (J2000), $10\farcm6$ from the position of NGC~6441 (shown in Fig.~\ref{6441_TSmap}). This position is consistent with that reported in~\citet{lat_13_GCs} and it likely lies outside the tidal radius of the cluster ($8\farcm00$) but it is still possible that some emission is from within the tidal radius given the current statistical and systematic errors. We obtained a single power-law fit with an index $2.4\pm0.1$, in which the integrated $\gamma$-ray flux is $F_{0.1-100\,\mathrm{GeV}}=(4.1\pm1.0)\times10^{-8}\mathrm{cm}^{-2}\,\mathrm{s}^{-1}$. This fit gives a TS value of 73. The photon spectrum can better be fit by a PLE where the photon index, $\Gamma$, is $0.4\pm0.9$ and the cutoff energy is (1.0$\pm$0.5)~GeV, giving $TS=100$. To further constrain the spectral parameters, we proceeded to fix $\Gamma$ at 0.2 to 2.0 (with steps of 0.2) while letting the normalization and $E_\mathrm{c}$ free, and found that TS$=$101 for $\Gamma_\gamma=$0.2, 0.4, and 0.6 and the TS value decreases with increasing $\Gamma$. If one fixes $\Gamma_\gamma$ at 0.4, $E_\mathrm{c}=(1.0\pm0.1)$~GeV is obtained. In this case, the integrated $\gamma$-ray flux is $F_{0.1-100\,\mathrm{GeV}}=(1.0\pm0.2)\times10^{-8}\mathrm{cm}^{-2}\,\mathrm{s}^{-1}$. The significance of the PLE model over the single power law model can be estimated by $TS_\mathrm{PLE}-TS_\mathrm{PL}=28$, which corresponds to 5.3$\sigma$. Therefore, we consider the PLE model to be a significantly better model than the power law model. The PLE model spectrum, as well as the data binned in different energies, is depicted in Fig.~\ref{ngc6441_spec}.
At a distance of 11.7~kpc~\citep{Valenti04}, the $\gamma$-ray luminosity of NGC~6441 (using the power law model) is $L_{0.1-100\,\mathrm{GeV}}=(3.7\pm0.9)\times10^{35}\mathrm{erg}\,\mathrm{s}^{-1}$, or $L_{0.1-100\,\mathrm{GeV}}=(1.3\pm0.2)\times10^{35}\mathrm{erg}\,\mathrm{s}^{-1}$ (using the PLE model).

\section{discussion}

\begin{figure*}
    \epsscale{1.}
    \plotone{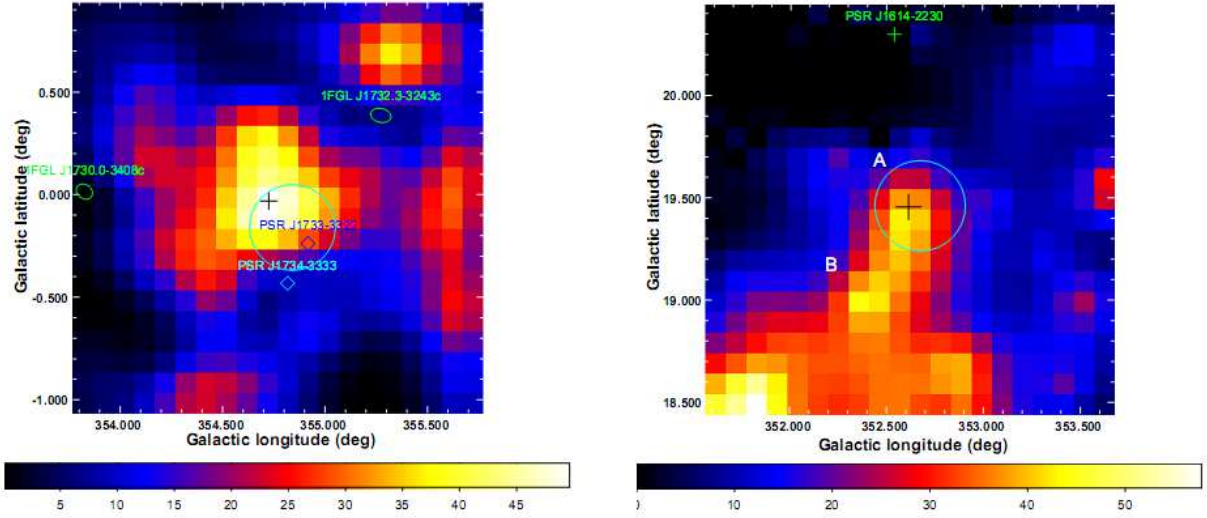}
      \caption{The Test-statistic (TS) maps of regions of 2$\degr\times$2$\degr$ centered on the best-fit centroids (labeled by crosses) of the $\gamma$-ray emission from Liller~1 (\emph{left}) and M80 (\emph{right}). The size of each cross indicates the 1-$\sigma$ statistical error in the determination of the centroid position. The color scale shows the TS value of every bin of an area $0\fdg1\times0\fdg1$. The light blue circles represent the tidal radius of the respective GCs compiled in~\citet[][2003 version]{harris_catalog}. \emph{Left}: PSR~J1733-3322 and PSR~J1734-3333, are marked by two white diamonds. The 68\% error positions of the LAT first catalog sources 1FGL~J1732.3-3243c and 1FGL~J1730.0-3408c (treated as background) are shown as eclipses. \emph{Right}: The position of the $\gamma$-ray MSP PSR J1614-2230~\citep{lat_8_field_GCs} is shown as a cross.}
         \label{Liller1_M80_TSmap}
   \end{figure*}

\begin{figure*}
    \epsscale{1.}
    \plotone{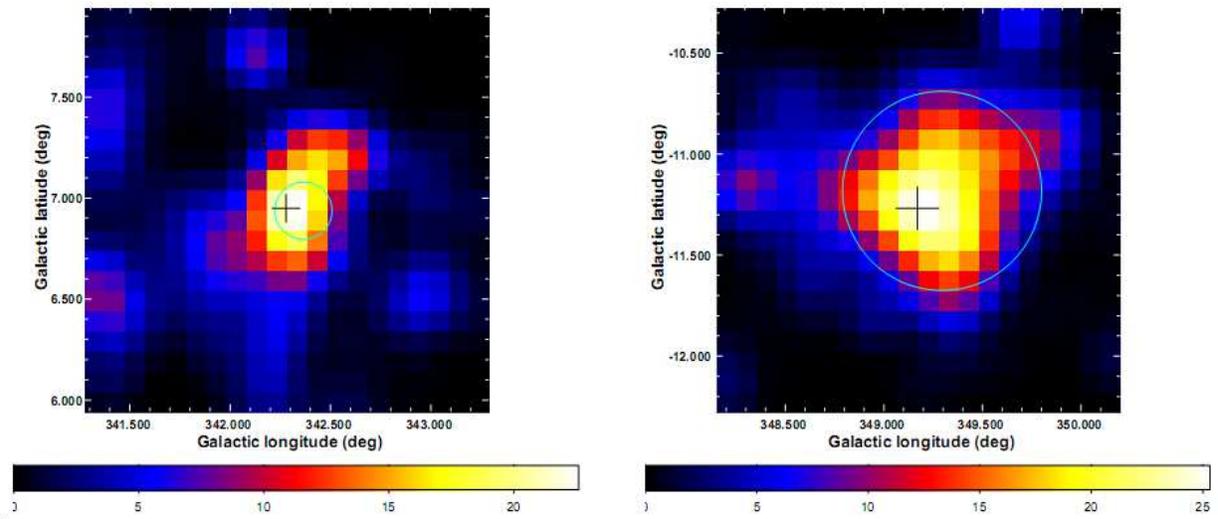}
      \caption{The TS maps of regions of 2$\degr\times$2$\degr$ centered on the best-fit centroids (labeled by crosses) of the $\gamma$-ray emission from NGC~6139 (\emph{left}) and NGC~6541 (\emph{right}). The circles represent the tidal radius of NGC~6139 (\emph{left}) and NGC~6541 (\emph{right})~\citep[][2003 version]{harris_catalog}. The meanings of the color scale and the size of the crosses are the same as in Figure~\ref{Liller1_M80_TSmap}.}
         \label{6139_6541_TSmap}
   \end{figure*}

\begin{figure*}
    \epsscale{1.}
    \plotone{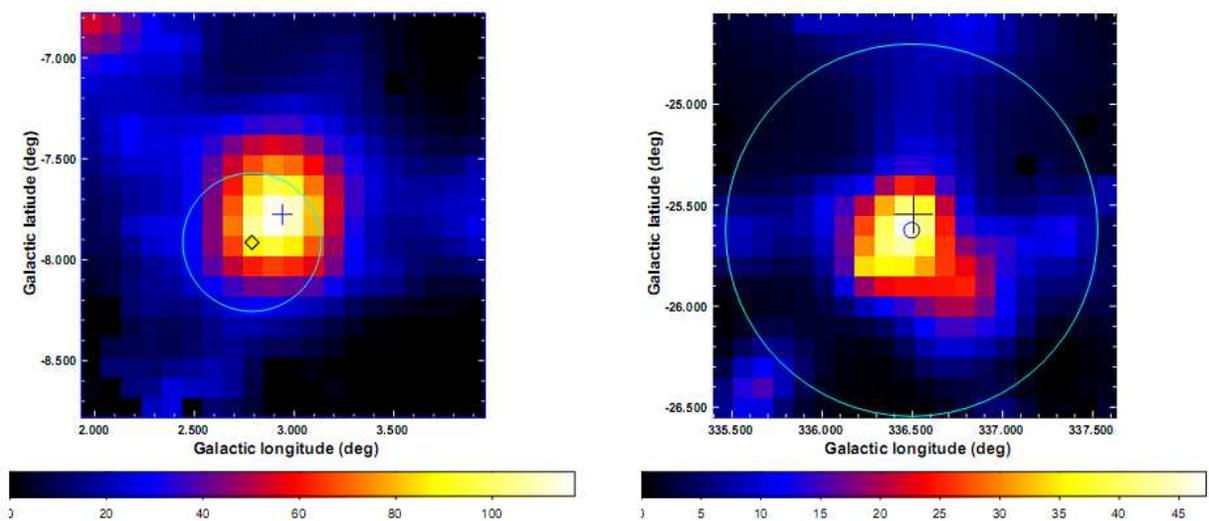}
      \caption{The TS maps of regions of 2$\degr\times$2$\degr$ centered on the best-fit centroids (labeled by crosses) of the $\gamma$-ray emission from NGC~6624 (\emph{left}) and NGC~6752 (\emph{right}). The TS map of a region of 2$\degr\times$2$\degr$ centered on the best-fit centroid (labeled by a cross) of the $\gamma$-ray emission next to NGC~6624. \emph{Left panel}: The circle represents the tidal radius of NGC~6624~\citep[][2003 version]{harris_catalog}. The white diamond shows the position of both PSR~B1820-30A and PSR~B1820-30B.  \emph{Right panel}: The outer and inner circle represents the tidal radius and half-mass radius of NGC~6752, respectively~\citep[][2003 version]{harris_catalog}. The meanings of the color scale and the size of the crosses are the same as in Figure~\ref{Liller1_M80_TSmap}.}
         \label{6624_6752_TSmap}
   \end{figure*}

\begin{figure}
    \epsscale{0.5}
    \plotone{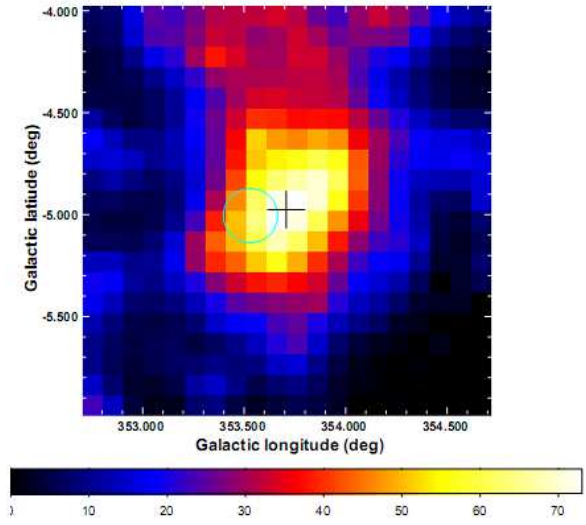}
      \caption{The TS map of a region of 2$\degr\times$2$\degr$ centered on the best-fit centroid (labeled by a cross) of the $\gamma$-ray emission next to NGC~6441. The circle represents the tidal radius of NGC~6441~\citep[][2003 version]{harris_catalog}. The meanings of the color scale and the size of the cross are the same as in Figure~\ref{Liller1_M80_TSmap}.}
         \label{6441_TSmap}
   \end{figure}

In this work, we report the discovery of three new $\gamma$-ray emitting GCs: Liller~1, NGC~6624, and NGC~6752, and we found evidence for $\gamma$-ray emission from M80, NGC~6139, and NGC~6541. We also confirmed the $\gamma$-ray detection next to NGC~6441 first mentioned by~\citet{lat_13_GCs}.
We searched through various X-ray and radio catalogs for plausible $\gamma$-ray sources in the neighborhood of the GCs. We found two pulsars next to Liller~1, as shown in Fig.~\ref{Liller1_M80_TSmap}~(left panel). The spin-down luminosity of PSR~J1733-3322 and PSR~J1734-3333, is $8.3\times10^{31}\mathrm{erg}\,\mathrm{s}^{-1}$ and $5.6\times10^{34}\mathrm{erg}\,\mathrm{s}^{-1}$, respectively \citep{Morris_Parks_psr}, i.e. more than an order of magnitude lower than the observed $\gamma$-ray luminosity of $10^{36}\mathrm{erg}\,\mathrm{s}^{-1}$. Together with their large offsets from the $\gamma$-ray emitting region, the association of the observed $\gamma$-rays with either pulsar is unlikely. In all cases, the GCs seem to be the only plausible counterparts of the observed $\gamma$-rays (we refer to region~A in the case of M80). Without clear counterparts in other wavelengths, the nature of the $\gamma$-ray emitting region \emph{B} shown in Fig.~\ref{Liller1_M80_TSmap} (right panel) remains unclear, and its relation with M80 cannot be confirmed or ruled out at this stage.

Several MSPs are known in NGC~6441, NGC~6624, and NGC~6752 (all located close to its core) but none has yet been uncovered in other cases~\citep[e.g.,][]{Hui_msp_stat}. We speculate that there is a population of MSPs -- the only known steady $\gamma$-ray emitters in GCs -- in the $\gamma$-ray emitting GCs that we have reported here. For example, the brighter X-ray population of M80 is in many ways similar to 47~Tucanae, the first $\gamma$-ray GC known, while the fainter X-ray sources may differ~\citep{Heinke03_M80_xray}. The discovery of $\gamma$-rays from M80, if confirmed, would suggest that it also hosts a population of MSPs. Under the assumption that the observed $\gamma$-ray luminosity depends solely on the number of MSPs (e.g. in the magnetospheric model), the predicted number of MSPs in the cluster using Eq.(1) of \citet{lat_13_GCs} is listed in Table~\ref{GC_results}. Future multiwavelength observations of any MSPs should help in understanding the $\gamma$-ray production mechanisms and the underlying MSP populations.

The $\gamma$-ray spectra are in general consistent with single power law model, and shows no convincing evidence for cut-off at high energies. It is mainly due to limited photon statistics. However, the cases of the more significant cases of Liller~1 and NGC~6624, in which $\gamma$-ray are found within the tidal radius, suggests that $>$10GeV emission exists at least for these two cases. If this is true, this high energy emission conform with the IC model, rather than the magnetospheric model. The $>$10GeV emission may correspond to the IC peak as predicted in~\citet{Bednarek07} and \citet{Cheng_ic10}.

If the $\gamma$-ray emission is associated with the GCs, one striking feature is the displacement of the $\gamma$-ray emission from the GC cores. We briefly discuss this feature in the context of two radiation scenarios commonly discussed in the literature.

In the pulsar magnetosphere model, one expects the emission to coincide spatially with the MSP population. While MSPs should concentrate in the GC core in general, the finite number of MSPs in a GC (i.e., $\sim$10, see Table~\ref{GC_results}, except for Liller~1) may invoke some offset of the MSPs -- and thus the $\gamma$-rays they emit -- from the core. Furthermore, in addition to the MSP population formed in the core, some MSPs are formed near the tidal radius of the GCs. The superposition of $\gamma$-rays from these two populations may not coincide with the GC core, but is at a certain distance from the core within the tidal radius. However, this scenario cannot accommodate the case of the $\gamma$-ray source next to NGC~6441.

In the inverse Compton models, the $\gamma$-ray emitting region depends not only on the distribution of the underlying MSPs, but also on the cooling timescale and the diffusion timescale of the accelerated particles. When the diffusion timescale is much smaller than the cooling timescale, the $\gamma-$rays may come from the outskirts of the GCs.

\begin{figure}
    \epsscale{1.2}
    \plotone{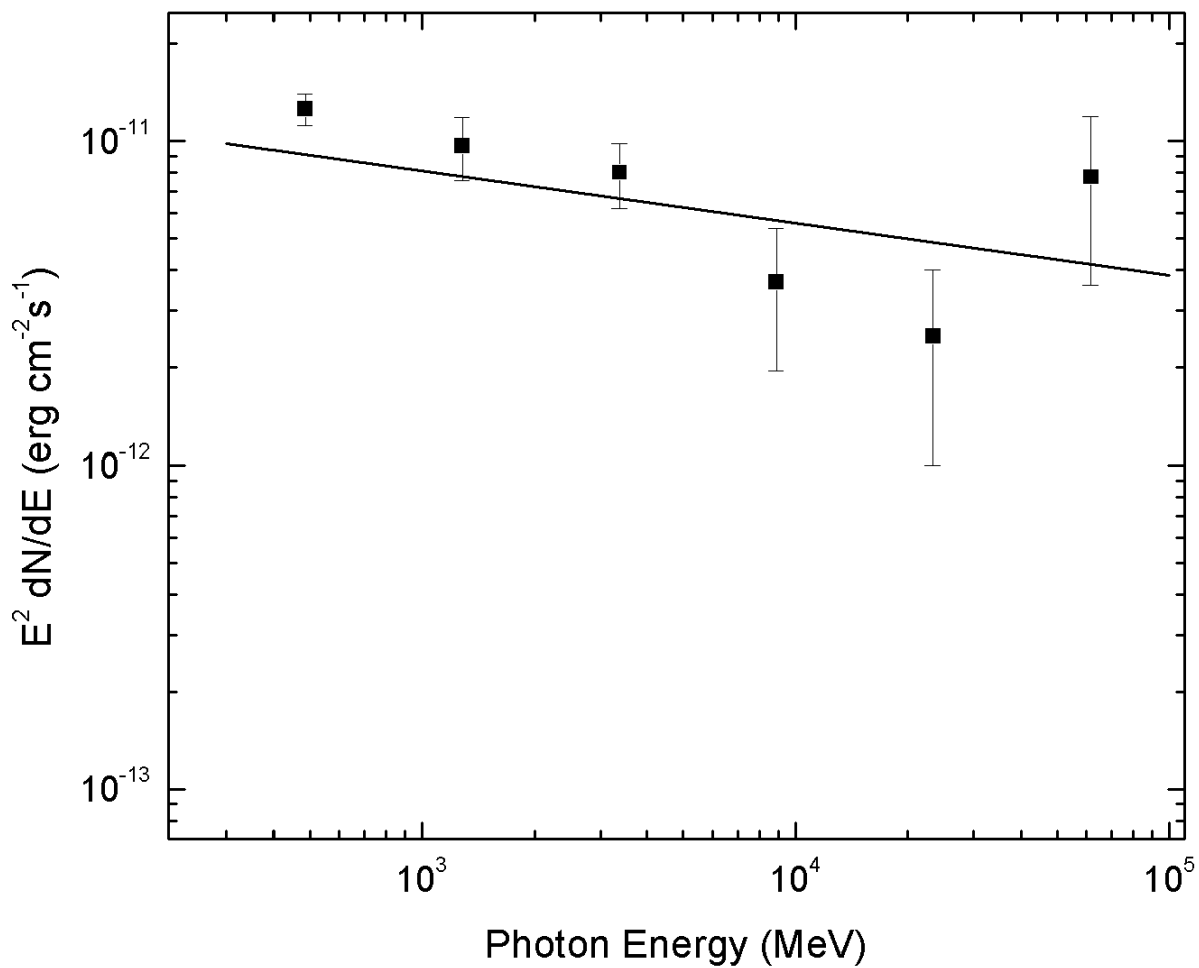}
      \caption{The spectrum ($E^2\times dN/dE$) of the $\gamma$-ray emission found within the tidal radius of Liller~1. Photons between 300~MeV and 100~GeV are divided into 6 logarithmically-equally spaced energy bins and The derived flux for each bin is plotted at the corresponding mean energy. The solid line indicates the best-fit single power law model. The rightmost point has a TS value of 16, corresponding to a significance level of four.}
         \label{Liller1_spec}
   \end{figure}

\begin{figure}
    \epsscale{1.2}
    \plotone{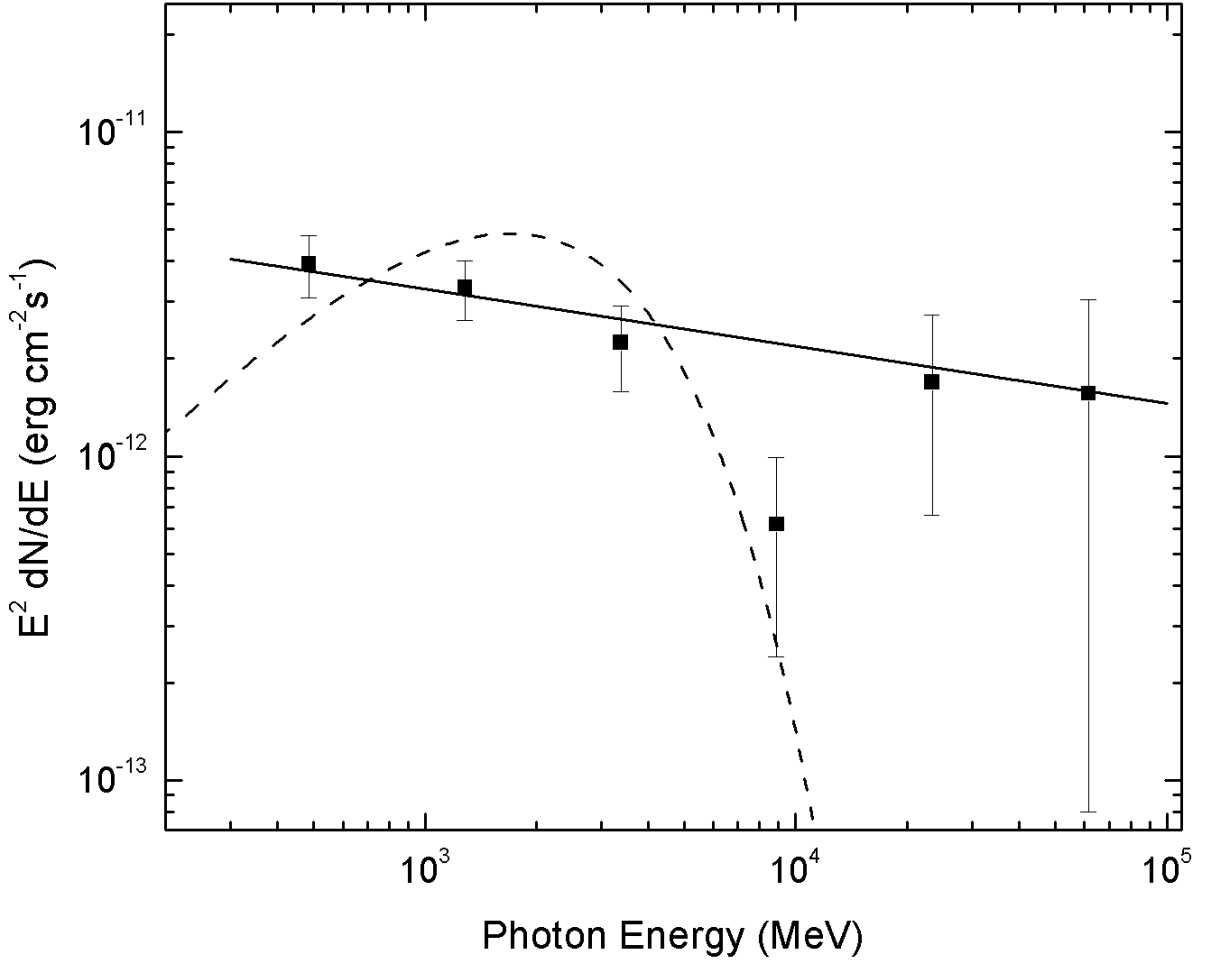}
      \caption{The spectrum ($E^2\times dN/dE$) of the $\gamma$-ray emission found within the tidal radius of NGC~6624. Photons between 300~MeV and 100~GeV are divided into 6 logarithmically-equally spaced energy bins and The derived flux for each bin is plotted at the corresponding mean energy. The solid line and the dashed line indicate the best-fit single power law model and the PLE model with $E_\mathrm{c}=1.5$~GeV, respectively. Each of the rightmost two points has a TS value of $\sim$10, corresponding to a significance level of about three each.}
         \label{ngc6624_spec}
   \end{figure}

In the cases of Liller 1 and NGC 6624, the
gamma-ray spectrum (cf. Fig.~\ref{Liller1_spec} and Fig.~\ref{ngc6624_spec}) can extend up to energies $>$40~GeV, which cannot be explained in terms of curvature radiation inside the light cylinder. Recently,
\citet{Cheng_ic10} have suggested that inverse Compton scattering between
background relativistic electrons/positrons in pulsar wind and the soft
photons from the galactic disk can produce the observed gamma-rays from
globular clusters. They predict that if the inverse Compton upscattered
Galactic infrared photons are responsible for GeV photons, then the Compton
upscattered galactic optical photons can produce gamma-rays up to 100~GeV.
They also predict that the gamma-rays are diffuse and are emitted
from region much beyond the core of the globular cluster. If this is true
the center of gamma-ray emission regions is affected by three factors:
the proper motion of the globular cluster, the asymmetric diffusion
coefficient and nearby optical/IR external sources like stars and nearby regions containing an enhanced amount of dust.

The discovery of several more $\gamma$-ray emitting GCs reported in this work has allowed population studies to be carried out in order to better understand the relationship between the $\gamma$-ray properties of GCs and other parameters of GCs.
\citet{lat_13_GCs} found that the stellar encounter rate scales with the $\gamma$-ray luminosity, based on a sample of 8 $\gamma$-ray emitting GCs. While the $\gamma$-ray luminosity should correlate with the number of MSPs in the clusters in both models, it is expected that the low-energy photon energy density is also a good estimator of $\gamma$-rays in the IC models. Recently, using a total of 15 GCs that are $\gamma$-ray sources or source candidates, including those reported in this work and those previously known~\citep{lat_13_GCs}, \citet{Hui11_correlation} have shown that correlations also exist between $\gamma$-ray luminosity and metallicity, as well as $\gamma$-ray luminosity and optical/infrared photon energy density.

To conclude, we report the detection of new $\gamma$-ray emitting GCs, based on a ranking list that replies on the encounter rate of the GCs. The $\gamma$-ray luminosity of Liller~1 is the highest of all known $\gamma$-ray GCs. Its non-detection before this work may be due to its position that lies on the Galactic plane, where Galactic diffuse emission is very strong. Liller~1 has very high encounter rate and metallicity, and may indeed host a large population of MSPs that have not been uncovered. Some $\gamma$-ray emitting GCs presented here are at distances of $\sim$10~kpc away from us, whereas \citet{lat_1st_cat} present the upper limits of known GCs out to a distance of 6~kpc only. Therefore, we have demonstrated that the study here has gone a further step towards finding new sources even at larger distances. This then helps to expand the sample of known $\gamma$-ray GCs for further studies.

\acknowledgments We acknowledge the use of data and software facilities from the FSSC, managed by the HEASARC at the Goddard Space Flight Center. CYH is supported by research fund of Chungnam National University in 2010. KSC is supported by a GRF grant of Hong Kong Government under HKU700908P, and AKHK is supported partly by the National Science Council of the Republic of China (Taiwan) through grant NSC96-2112-M-007-037-MY3 and NSC99-2112-M-007-004-MY3.


\clearpage

\begin{landscape}
\begin{table}
\caption{Region-of-interest (ROI), minimum photon energy ($E_\mathrm{min}$), and maximum photon energy ($E_\mathrm{max}$) in the analysis of the GCs} \label{GC_data}
\centering
\begin{tabular}{l@{}cc@{}ccc}
\hline\hline
GC name  & \multicolumn{2}{c}{ROI center} & ROI radius & $E_\mathrm{min}$ & $E_\mathrm{max}$ \\
         & R.A. (J2000) & Dec. (J2000) & & &  \\
         & ($\degr$) & ($\degr$) & ($\degr$)    & (GeV)     & (GeV)  \\
\hline
Liller~1 & $263.20$       & $-33.39$              & 5            & 0.3              & 100    \\
M80      & $244.23$       & $-23.02$              & 10           & 0.2              & 100    \\
NGC~6139 & $246.83$       & $-38.90$              & 6            & 0.3              & 100    \\
NGC~6441 & $267.63$       & $-36.89$              & 5            & 0.3              & 100    \\
NGC~6541 & $272.04$       & $-43.85$              & 6            & 0.3              & 100    \\
NGC~6624 & $275.86$       & $-30.16$              & 6            & 0.3              & 100    \\
NGC~6752 & $287.57$       & $-59.96$              & 10           & 0.2              & 100    \\
\hline
\end{tabular}
\end{table}

\begin{table}
\tabletypesize{\scriptsize}
\caption{Summary of $\gamma$-ray results of the GCs} \label{GC_results}
\begin{tabular}{llr@{}rrcc@{}cccc@{}c}
\hline\hline
GC name  & Spectral & TS    & Significance & $N_\mathrm{ph}$\tablenotemark{a} & Photon index\tablenotemark{b} & Cutoff energy & Photon flux\tablenotemark{c} & Energy flux\tablenotemark{d} & Distance\tablenotemark{e} & Luminosity\tablenotemark{f} & $N_\mathrm{MSP}$\tablenotemark{g} \\
         &  model   &       & ($\sigma$)   &                                  & ($\Gamma$) & ($E_\mathrm{c}$, GeV) &  & & (kpc) &  & \\
\hline
Liller~1 & PL  & 107 & 10.3 & 966 & $2.2\pm0.1$ & \nodata & $69\pm23$   & $53\pm18$   & 9.6 & $59\pm20$ & $410^{+480}_{-210}$\\
M80      & PL  & 27  & 5.2  & 125 & $2.0\pm0.2$ & \nodata & $5.5\pm3.6$ & $6.6\pm4.2$ & 10.3$^{+0.8}_{-0.7}$ & 8.4$^{+7.6}_{-5.8}$ & $58^{+123}_{-45}$ \\
NGC~6139 & PL  & 32  & 5.6  & 154 & $2.1\pm0.2$ & \nodata & $9.9\pm5.4$ & $8.8\pm4.8$ & 10.1 & $11\pm6$ &$75^{+114}_{-50}$\\
NGC~6441 & PL  & 73  & 8.5  & 451 & $2.4\pm0.1$ & \nodata & $41\pm10$   & $23\pm6$    & 11.7 & $37\pm9$ & \nodata\tablenotemark{h} \\
         & PLE & 101 & 10.0 & 380 & 0.4\tablenotemark{i}& $1.0\pm0.1$ & $10\pm2$ & $8.1\pm1.0$ & 11.7 & $13\pm2$ &\nodata\\
NGC~6541 & PL  & 19  & 4.4  & 109 & $2.2\pm0.2$ & \nodata & $9.5\pm4.4$ & $6.5\pm3.0$ & $6.9\pm0.7$ & $3.7^{+2.9}_{-2.1}$ & $26^{+49}_{-18}$\\
NGC~6624 & PL  & 121 & 11.0 & 362 & $2.2\pm0.1$ & \nodata & $26\pm6$    & $20\pm5$    & 7.9& $15\pm3$ & $103^{+104}_{-46}$\\
         & PLE & 121 & 11.0 & 326 & $0.9\pm0.2$ & 1.5   & $10\pm3$    & $7.4\pm1.0$ & 7.9  & $5.5\pm0.7$ & \nodata \\
NGC~6752 & PL  & 49  & 7.0  & 130 & $2.0\pm0.2$ & \nodata & $6.0\pm2.8$ & $6.0\pm2.8$ & $4.4\pm0.1$ & $1.4\pm0.7$ & $10^{+15}_{-6}$\\
\hline
\tablenotetext{a}{~Modeled photon number associated with the corresponding $\gamma$-ray source in the best likelihood fit}
\tablenotetext{b}{~All the quoted errors are statistical and $1\sigma$ for one parameter of interest.}
\tablenotetext{c}{~Integrated 0.1--100~GeV photon flux in unit of 10$^{-9}$cm$^{-2}$s$^{-1}$}
\tablenotetext{d}{~Integrated 0.1--100~GeV energy flux in unit of 10$^{-12}$erg~cm$^{-2}$s$^{-1}$}
\tablenotetext{e}{~Distance adopted from~\citet[][2003 version]{harris_catalog}, except for M80~\citep{brocato_M80_dist}, NGC~6541~\citep{LeeCarney06}, and \\ NGC~6441~\citep{Valenti04}.}
\tablenotetext{f}{~0.1--100~GeV luminosity in unit of 10$^{34}$erg~s$^{-1}$}
\tablenotetext{g}{~Predicted number of MSPs in the cluster using Eq.(1) of \citet{lat_13_GCs}, under the assumption that the observed $\gamma$-ray luminosity \\ depends solely on the number of MSPs}
\tablenotetext{h}{~In this case, it is unlikely that the number of MSPs is the sole factor in determining the observed $\gamma$-ray luminosity due to the offset}
\tablenotetext{i}{~Model parameters without quoted errors are fixed at the value given.}
\end{tabular}
\end{table}
\clearpage
\end{landscape}

\begin{thebibliography}{}
\bibitem[Abdo et al.(2009a)]{lat_47Tuc_Science} Abdo, A. A., et al. (Fermi/LAT Collaboration)\ 2009a, Science, 325, 845
\bibitem[Abdo et al.(2009b)]{lat_8_field_GCs} Abdo, A.~A., et al. (Fermi/LAT Collaboration)\ 2009b, Science, 325, 848
\bibitem[Abdo et al.(2009c)]{bsl_lat} Abdo, A. A., Ackermann, M., Ajello, M., et al. (Fermi/LAT Collaboration)\ 2009c, \apjs, 183, 46
\bibitem[Abdo et al.(2010a)]{lat_1st_psr_cat} Abdo, A. A., et al. (Fermi/LAT Collaboration)\ 2010a, \apjs, 187, 460
\bibitem[Abdo et al.(2010b)]{lat_1st_cat} Abdo, A. A., et al. (Fermi/LAT Collaboration)\ 2010b, \apjs, 188, 405
\bibitem[Abdo et al.(2010c)]{lat_13_GCs} Abdo, A. A., et al. (Fermi/LAT Collaboration)\ 2010c, \aap, 524, 75
\bibitem[Aharonian et al.(2009)]{hess_47Tuc} Aharonian, F., et al. (HESS Collaboration)\ 2009, \aap, 499, 273
\bibitem[Anderhub et al.(2009)]{magic_M13} Anderhub, H., et al. (MAGIC Collaboration)\ 2009, \apj, 702, 266
\bibitem[Atwood et al.(2009)]{lat_technical} Atwood, W.~B., et al. (Fermi/LAT Collaboration)\ 2009, \apj, 697, 1071
\bibitem[Bednarek \& Sitarek(2007)]{Bednarek07} Bednarek, W., \& Sitarek, J.\ 2007, \mnras, 377, 920
\bibitem[Brocato et al.(1998)]{brocato_M80_dist} Brocato, E., Castellani, V., Scotti, G.~A., Saviane, I., Piotto, G., \& Ferraro, F.~R.\ 1998, \aap, 335, 929
\bibitem[Cheng et al.(2010)]{Cheng_ic10} Cheng, K.~S., Chernyshov, D.~O., Dogiel, V.~A., Hui, C.~Y., \& Kong, A.~K.~H.\ 2010, \apj, 723, 1219
\bibitem[Gendre et al.(2003)]{Gendre2003} Gendre, B., Barret, D., \& Webb, N.\ 2003, \aap, 403, L11
\bibitem[Gnedin et al.(2002)]{Gnedin02} Gnedin, O. Y., Zhao, H. S., Pringle, J. E., Fall, S. M., Livio, M., \& Meylan, G.\ 2002, \apj, 568, L23
\bibitem[Harris(1996)]{harris_catalog} Harris, W.~E. 1996, \aj, 112, 1487 (2003 version)
\bibitem[Heinke et al.(2003)]{Heinke03_M80_xray} Heinke, C.~O., Grindlay, J.~E., Edmonds, P.~D., Lloyd, D.~A., Murray, S.~S., Cohn, H.~N., \& Lugger, P.~M.\ 2003, \apj, 598, 516
\bibitem[Hui et al.(2009)]{Hui_msp_stat} Hui, C.~Y., Huang, H.~H., Cheng, K.~S., Taam, R.~E.,
\& Becker, W.\ 2009, Astronomical Society of the Pacific Conference Series, 404, 149
\bibitem[Hui et al.(2010)]{Hui10_metallicity} Hui, C.~Y., Cheng, K.~S., \& Taam, R.~E.\ 2010, \apj, 714, 1149
\bibitem[Hui et al.(2011)]{Hui11_correlation} Hui, C.~Y., Cheng, K.~S., Wang, Y., Tam, P.~H.~T., Kong, A.~K.~H., Chernyshov, D.~O., \& Dogiel, V.~A.\ 2011, \apj, 726, 100
\bibitem[Kong et al.(2010)]{Kong10_Terzan5} Kong, A.~K.~H., Hui, C.~Y., \& Cheng, K.~S.\ 2010, \apjl, 712, L36
\bibitem[Lee \& Carney(2006)]{LeeCarney06} Lee, J.~W. \& Carney, B.~W.\ 2006, \aj, 132, 2171
\bibitem[Manchester et al.(2005)]{atnf_cat} Manchester, R.~N., Hobbs, G.~B., Teoh, A., \& Hobbs, M.\ 2005, \aj, 129, 1993
\bibitem[Mattox et al.(1996)]{Mattox_96} Mattox, J.~R., et al.\ 1996, \apj, 461, 396
\bibitem[Morris et al.(2002)]{Morris_Parks_psr} Morris, D.~J., et al.\ 2002, \mnras, 335, 275
\bibitem[Ortolani et al.(2007)]{Ortolani_Liller1_UKS1_dist} Ortolani, S., Barbuy, B., Bica, E., Zoccali, M., \& Renzini, A.\ 2007, \aap, 470, 1043
\bibitem[Pooley et al.(2003)]{Pooley03} Pooley, D., et al.\ 2003, \apjl, 591, L131
\bibitem[Pooley \& Hut(2006)]{Pooley06} Pooley, D., \& Hut, P.\ 2006, \apjl, 646, L143
\bibitem[Valenti et al.(2004)]{Valenti04} Valenti, E., Ferraro, F.~R., \& Origlia, L.\ 2004, \mnras, 351, 1204
\bibitem[Venter \& de Jager(2008)]{Venter08} Venter, C., \& de Jager, O.~C.\ 2008, \apjl, 680, L125
\bibitem[Verbunt \& Hut(1987)]{Verbunt87} Verbunt, F., \& Hut, P.\ 1987, Proceedings of the IAU Symposium No. 125, The Origin and Evolution of Neutron Stars, 187

\end{thebibliography}
\end{document}